\documentclass[final,5p,times,a4paper,sort&compress]{elsarticle}
\usepackage{graphicx}
\usepackage[usenames]{color}
\usepackage{amsmath}
\usepackage{amssymb}
\usepackage{polymers}
\usepackage{sizeredc}
\usepackage{color,soul}

\begin{document}
\begin{frontmatter}

\title{{  Effects of Spatial Curvature on Blackbody Radiation: Modifications to Energy Distribution and Fundamental Laws}}

\author[l1]{Somayeh Kourkinejat}
\ead{s.koorkinejat@sci.ui.ac.ir}
\author[l1,l2]{Ali Mahdifar}
\ead{a.mahdifar@sci.ui.as.ir}
\author[l3,l4]{Ehsan Amooghorban}
\ead{ehsan.amooghorban@sku.ac.ir}
\address[l1]{Physics Department, University of Isfahan, Hezar Jerib St. Isfahan, 81764-73441, Iran.}
\address[l2]{Quantum Optics Group, Department of Physics, University of Isfahan, Hezar Jerib St. Isfahan, 81746-73441, Isfahan, Iran.}
\address[l3]{Department of Physics, Faculty of Science, Shahrekord University P. O. Box 115, Shahrekord, Iran}
\address[l4]{Nanotechnology Research Center, Shahrekord University, 8818634141, Shahrekord, Iran}
\cortext[cor1]{Corresponding author}
\begin{abstract}
In this paper, we investigate the effects of spatial curvature on blackbody radiation. By employing an analog model of general relativity, we replace the conventional straight-line harmonic oscillators used to model blackbody radiation with oscillators on a circle. This innovative approach provides an effective framework for describing blackbody radiation influenced by spatial curvature.
We derive the curvature-dependent Planck energy distribution and find that moving from flat to curved space results in a reduction in both the height and width of the Planck function. Moreover, increasing the curvature leads to a pronounced redshift in the peak frequency. We also analyze the influence of spatial curvature on the Stefan-Boltzmann law, Rayleigh-Jeans law, and Wien law.
\end{abstract}
\end{frontmatter}
\section{Introduction}\label{sec1}
Blackbody radiation has played a fundamental role in the development of modern physics \cite{Giliberti2024old}.
It has contributed greatly to the development of various scientific disciplines, including quantum mechanics, quantum statistical physics and stellar astrophysics~\cite{B1,guha2017q}.
Historically, blackbody radiation has served as a cornerstone for testing advancements in statistical physics, owing to its capacity to describe experimentally realizable phenomena.
Motivated by the need to explain early measurements of light spectra and heat radiation from heated bodies, Planck derived his expression for blackbody radiation by postulating the quantization of energy states in a simple harmonic oscillator \cite{planck1914theory,blomstedt2015blackbody}.
This work was later refined by contributions from Einstein, Bose, and Pauli, among others~\cite{ourabah2014planck,Kramm2009PlancksBR}.
The evolution of modern physics witnessed the emergence of key principles governing blackbody radiation, including Planck’s formula, Wien’s displacement law, and Stefan’s law \cite{guo2023planck}. The generalized Planck radiation law has been very useful in interpreting data from the cosmic microwave background~\cite{tsallis1995generalization}.


In stellar astrophysics, the effective surface temperature of a star, such as our Sun, can be estimated by analyzing its radiation spectrum. By modeling the star as an ideal blackbody and comparing its observed spectrum to that of a typical blackbody in thermal equilibrium, astronomers can determine the star's effective temperature~\cite{bohm1989introduction}.
However, the spectrum of an ideal blackbody is an abstraction. General relativity dictates that the curvature of spacetime, caused by a star's mass, significantly influences its radiation spectrum. This implies that no true ideal blackbody spectrum, unaffected by curvature, exists in nature. To accurately estimate the surface temperature of a star, it is essential to account for the influence of the curvature on the blackbody radiation spectrum.

Einstein's theory of general relativity establishes that massive objects curve the surrounding spacetime, producing gravitational effects inherently linked to this curvature \cite{R47, R54}. Due to the often subtle nature of gravitational effects, some predictions of general relativity are challenging to verify experimentally. To address this, researchers have proposed alternative analog models to investigate the effects of general relativity on various phenomena using laboratory setups~\cite{Xu:21,R30,Xu:18,R14,R39,R33,R29,R48,R49,schutzhold2002gravity,Tavakoli18}.
For instance, analog models have been used to explore black hole physics, aiding in the development of thermodynamic theories for these objects and the discovery of effects like Hawking radiation~\cite{R16}.
%
%
%
In~\cite{R19}, an alternative approach to understanding spatial curvature is proposed. This approach simplifies the problem by treating time as a constant and reducing the spatial dimensions to one or two, thereby enabling the examination of systems on curved surfaces such as spheres or circles.

%

In this paper, our goal is to investigate the effects of spatial curvature on blackbody radiation. To do so, we adopt an analog model of general relativity, utilizing oscillators on a circle with radius $R$ (and spatial curvature $\Lambda = 1/R^{2}$), rather than the conventional harmonic oscillators on a straight line. This approach allows us to describe curvature-affected blackbody radiation effectively. Since the eigenenergies of these oscillators depend on the curvature of the circle, our model serves as a powerful analog model to explore how the curvature of physical space influences the radiation properties of a blackbody.

This paper is organized as follows: In Sec.~\ref{Quantum Harmonic Oscillator},  we present our analog model as a simple quantum harmonic oscillator on a circle.
Sec.~\ref{spatial curvature in black-body}, is devoted to obtain curvature-dependent partition function and blackbody radiation formula of our analog model. Generalized curvature-dependent Planck’s energy density distribution,  Stefan-Boltzmann law, generalized Rayleigh-Jeans law, and generalized Wien law are given in Sec.~\ref{Planck distribution}.  We study  generalized curvature-dependent Wien’s displacement relationship  for our generalized blackbody radiation in Sec.~\ref{Wien law}. Finally, the summary and concluding remarks are given in Sec.~\ref{SUMMARY}.
%
\section{Quantum Harmonic Oscillator on Circle}\label{Quantum Harmonic Oscillator}
To investigate the impact of space curvature on blackbody radiation, we employ an analog model of general relativity. In this model, instead of using harmonic oscillators on a straight line to model radiation,
we utilize harmonic oscillators on a circle with radius $R$ (and spatial curvature $\Lambda = \frac{1}{R^{2}}$) to describe the curvature-affected black body radiation.
This section provides a brief introduction to the concept of quantum harmonic oscillators on a circle and sets the context for our analysis.

Recently, some of the authors of this paper explored the quantum dynamics of an oscillator with  $ m= 1$ on a circle of radius $R$. The Hamiltonian for this quantum harmonic oscillator was derived using the gnomonic projection~\cite{doi:10.1142/S0219887822501407}, as follows:
\begin{equation}\label{1}
 \hat{H} (\Lambda) = \frac{\hbar^{2}}{2} \Big[-(1+ \Lambda x^{2})^{2} \frac{d^{2}}{dx^{2}} - 2\Lambda x (1+ \Lambda x^{2})\frac{d}{dx}\Big] +\frac{\omega^{2}}{2} x^{2}.
\end{equation}
The energy eigenvalues of the quantum oscillator on a circle are obtained as follows:
\begin{equation}\label{2}
E_{n}(\Lambda) = \hbar\omega\Big[\Gamma (n + \frac{1}{2})+ \frac{\Lambda}{2}  n^{2}\Big],
\end{equation}
where
\begin{equation}\label{3}
\Gamma = \frac{ \Lambda + \sqrt{{\Lambda}^{2} + 4} }{2}.
\end{equation}
It is worth noting that in the limit of $\Lambda \rightarrow 0$, we obtain $\Gamma \rightarrow 1$, and Eq.~\eqref{2} reduces to the energy eigenvalues of the quantum harmonic oscillator on a straight line: $E_{n}(\Lambda = 0) =\hbar\omega (n + \frac{1}{2})$.
%
\section{ Effects of the spatial curvature in black-body radiation }~\label{spatial curvature in black-body}
A blackbody is an ideal object that absorbs all incident electromagnetic radiation, irrespective of wavelength, while simultaneously emitting thermal radiation. Blackbody radiation can be experimentally simulated using a cavity that maintains thermal equilibrium with the radiation inside it~\cite{bransden1989introduction,rybicki1991radiative}.

To investigate the impact of space curvature on blackbody radiation, we consider a radiation cavity with volume $V$ and temperature $T$. In our analog model, the curvature-affected radiation is investigated as an ensemble of quantum harmonic oscillators on a circle,  with eigen-energy $E_{n}(\Lambda)$,  in thermal equilibrium with one another and with the cavity walls at temperature $T$. Consequently, the radiation emitted through a small hole in one of the walls will exhibit the characteristics of an ideal blackbody.
We also assume that the oscillators are  distinguishable from each others. This assumption is justified because these oscillators  on a circle merely represent the energy levels available within our system. In other words, it is actually photons surrounding a massive stellar object, behaving as an ideal curvature-affected  blackbody, that distribute themselves over the various curvature–dependent oscillator levels.
Consequently, these oscillators are expected to obey Maxwell-Boltzmann statistics.

To calculate the partition function for a single oscillator on a circle, we compute:
\begin{equation}\label{4}
Q_{1} = \sum_{n}^{\infty} e^{-\beta E_{n}(\Lambda)},
\end{equation}
where $\beta = 1/kT$. By using the energy levels \eqref{2}, after some straightforward calculations, we obtain the  partition function as:
\begin{eqnarray}\label{5}
Q_{1} &=& \sum_{E} e^{-\hbar \omega \beta\Gamma(n+\frac{1}{2})-\frac{1}{2} \hbar\omega\beta \Lambda n^{2}}
 \\\
& = & \frac{1}{\sinh \frac{\beta \hbar \omega}{2}}\Big[1-\Lambda \frac{\beta \hbar \omega }{4} \coth^{2}\frac{\beta \hbar \omega }{2}\Big] + O(\Lambda^{2}),  \nonumber
\end{eqnarray}
where $O(\Lambda^{2})$ represents the terms of order equal to or greater than $\Lambda^{2}$, which are negligible for this system.
Using Eq.~\ref{5}, the partition function for a system of $N$ oscillators on a circle is given by
\begin{eqnarray}\label{6}
Q_{N} &=& \Big[\sum_{E} e^{-\hbar \omega \beta\Gamma(n+\frac{1}{2})-\frac{1}{2} \hbar\omega\beta \Lambda n^{2}}\Big]^{N}   \nonumber\\
&\simeq&  \Big[{\frac{1}{\sinh \frac{\beta \hbar \omega}{2}}1-\Lambda \frac{\beta \hbar \omega }{4} \coth^{2}\frac{\beta \hbar \omega }{2}\Big]}^{N}.
\end{eqnarray}
Note that the above equation is derived under the assumption of distinguishability between the oscillators.
%
\section{ curvature–dependent  Planck’s energy density distribution}\label{Planck distribution}
In this section, we examine the effects of curvature on various properties of blackbody radiation. The internal energy of this system  can be expressed in terms of the partition function~\eqref{6}  as follows:
\begin{eqnarray}\label{8}
U&=&-\frac{\partial \ln Q_{N}}{\partial \beta}\\
&=&N \hbar \omega \Big[\frac{\Gamma}{2} + \frac{\sum^{\infty}_{n=0}(\Gamma n + \frac{\Lambda}{2} n^{2})e^{-\beta \hbar \omega (\Gamma n + \frac{\Lambda}{2} n^{2})}}{\sum_{n}e^{-\beta \hbar \omega (\Gamma n +\frac{\Lambda}{2} n^{2})}}\Big]. \nonumber
\end{eqnarray}
The first term refers to the zero-point fluctuations $E_{n=0}(\Lambda)= \frac{\Gamma}{2}\hbar \omega $,  and the second term can be identified with thermal radiation actually  emitted by a blackbody at a temperature $T$, which depends on the spatial curvature.
In the following, we analyze the obtained curvature-dependent  blackbody radiation formula in the limit of high temperature $T$, low frequency $\omega$ and  low spatial curvature,  i.e., $\frac{\hbar \omega}{k T}\Lambda\ll 1$. Under these limiting conditions, we can write:
 \begin{eqnarray}\label{9}
 U &\simeq & \frac{N \hbar \omega }{2} \coth \frac{\beta \hbar \omega}{2} \\\
 &\times& \Big[1+\frac{\Lambda}{2} \frac{(1-e^{-2\beta \hbar \omega} - 4 \beta \hbar \omega e^{\beta \hbar \omega})}{(1-e^{-\beta \hbar \omega})^{2}}\Big]. \nonumber
 \end{eqnarray}
In the flat limit $\Lambda \rightarrow 0$, the above $\Lambda$-dependent formula reduces to the standard energy formula for a harmonic oscillator: $U=\frac{N \hbar \omega }{2} \coth \frac{\beta \hbar \omega}{2}$ \cite{beale1996statistical}.

Within our generalized $\Lambda$-dependent analog model, the average energy of a harmonic oscillator on a circle in thermal equilibrium, excluding the zero-point term $E_{n=0}(\Lambda)$, is expressed as:
\begin{equation}\label{10}
\langle \varepsilon \rangle = \frac{U- \frac{N\hbar\omega \Lambda}{2}}{N}=  \hbar\omega \frac{\sum_{n} (\Gamma n + \frac{\Lambda}{2} n^{2}) e^{-\beta \hbar \omega (\Gamma n + \frac{\Lambda}{2} n^{2})}}{\sum_{n} e^{-\beta \hbar \omega (\Gamma n + \frac{\Lambda}{2} n^{2})}}.
\end{equation}
Furthermore, the number of normal modes of vibration per unit volume of the cavity in the frequency range $(\omega, \omega + d\omega)$ is given by Rayleigh-Jeans expression~\cite{TIRNAKLI1997657,guo2023planck}:
\begin{equation}\label{11}
dN = 2 (4 \pi^{2} \nu^{2}) \frac{1}{c^{3}}d\nu = (\frac{\omega^{2}}{\pi^{2} c^{3}}) d\omega,
\end{equation}
where $ \omega=2 \pi \upsilon $, and  $c$  denotes the speed of light.
Thus, by using Eqs. \eqref{10} and \eqref{11}, the energy density of radiation between the frequency $(\omega, \omega + d \omega)$ can be calculated as follows:
 \begin{eqnarray}\label{12}
u(\omega, T, \Lambda)d\omega & =& \langle \varepsilon \rangle (\frac{\omega^{2}}{\pi^{2} c^{3}} d\omega) \\\
 &=&\frac{\hbar \omega^{3}}{\pi^{2} c^{3}} \frac{\sum_{n} (\Gamma n + \frac{\Lambda}{2} n^{2}) e^{-\beta \hbar \omega (\Gamma n + \frac{\Lambda}{2} n^{2})}}{\sum_{n} e^{-\beta \hbar \omega (\Gamma n + \frac{\Lambda}{2} n^{2})}}d\omega \nonumber  \\\
& \cong & \frac{\hbar \omega^{3}}{\pi^{2} c^{3}}  \frac{ 1}{e^{\beta \hbar \omega}-1}  \nonumber  \\\
&\times& \Big[1-\Lambda \frac{(\beta \hbar \omega +1) e^{-\beta \hbar \omega}+(\beta \hbar \omega-1)}{(1-e^{-\beta \hbar \omega})^{2}}\Big]d\omega  \nonumber .
\end{eqnarray}
The above energy density is the curvature-dependent Planck’s formula for the distribution of energy over the blackbody spectrum. As expected, in the $\Lambda \rightarrow 0$ limit, this expression reduces to the standard Planck’s formula \cite{beale1996statistical}.

\begin{figure}
\centering
 \includegraphics[width=0.8\columnwidth]{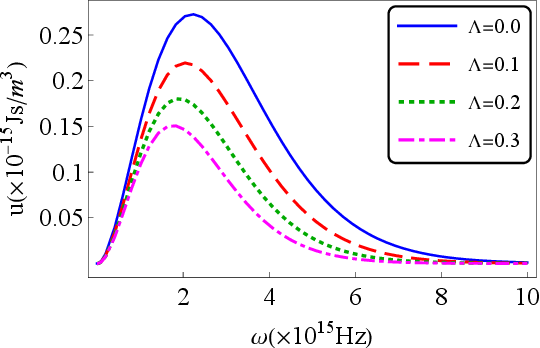}
\caption{  Energy density $u(\omega,T,\Lambda)$ versus $\omega$ for $T=6000K$; the thin-solid blue curve corresponds $\Lambda=0$, the dashed red curve to $\Lambda=0.1$, the dotted green curve to $\Lambda=0.2$, the dotted-dashed magenta curve to $\Lambda=0.3$.}
\label{fig:1}
\end{figure}

In Fig. \ref{fig:1} we have plotted the  energy density $u(\omega,T,\Lambda)$ of our curvature-dependent blackbody radiation as a function of frequency $\omega$ for different values of the spatial curvature, with the temperature fixed at
$T=6000K$, which corresponds to the estimated effective surface temperature of a Sun-like star.
As seen in this figure, an increase in spatial curvature results in a decrease in both the height and width of the Planck radiation function. Furthermore, a significant redshift in the peak frequency is evident with increasing curvature.
\begin{table}[b]
  \centering
   \caption{Value of the frequency $\omega_{m}$ at which   $u(\omega_{m})\equiv u_{max}$ reaches its  maximum  value, for different $\Lambda$}\label{tab1-1}
  \begin{tabular}{p{0.1\textwidth} p{0.2\textwidth} p{0.15\textwidth}}\hline
\hline
   $\Lambda$          &   $\omega_{m}(Hz)$   &             $u_{max}(J m^{-3} s)$ \\
  \hline
   $0.0$     &       $ 2.2713\times10^{15}$     &   $2.72515\times10^{-16}$  \\
   $0.1$     &       $2.09769\times10^{15}$     &    $2.19342\times10^{-16}$    \\
   $0.2$     &       $2.00282\times10^{15 }$    &    $1.79894\times10^{-16}$   \\
   $0.3$     &      $1.71581\times10^{15}$       &    $1.50799\times10^{-16}$    \\
  \hline
\hline
\end{tabular}
\end{table}
To get more insight in this curvature–dependent plot,  we can calculate the maximum value of the energy spectral distribution, by setting derivative of  Eq.~\eqref{12} to zero.
In Table \ref{tab1-1}, the obtained maximum values are shown, for different values of the parameter  $\Lambda $, with $T=6000K$.

\begin{figure}
\centering
  \includegraphics[width=0.8\columnwidth]{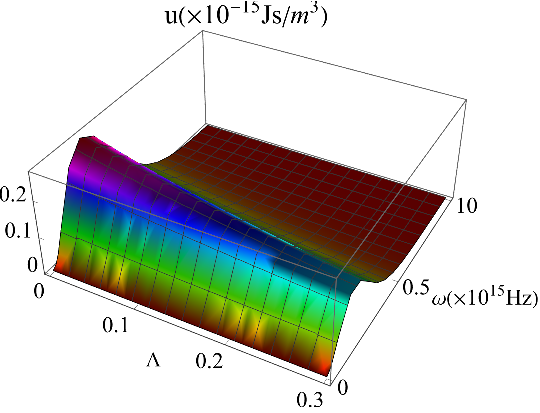}
\caption{ Energy density $u(\omega,T,\Lambda)$  versus $\omega$ and $\Lambda$ for $T=6000K$.}
 \label{fig:2}
\end{figure}

In order to illustrate the obtained results more clearly, in Fig. \ref{fig:2} we have also displayed a tree-dimensional plot of energy distribution versus the curvature $\Lambda$ and frequency $\omega$.

It is worth noting that the Hawking-Bekenstein calculations lead to the following relationship between black hole mass $M$ and it’s temperature $T$ \cite{weinstein2021demons}:
\begin{equation}\label{151}
T = \frac{c^{3}\hbar}{8 \pi G M k},
\end{equation}
where $G$ and $k$ are the universal gravitational and  the Boltzmann constants, respectively.
This relation indicates that a black hole can be considered as a blackbody, with its temperature inversely proportional to its mass. Thus, as the mass of a black hole increases, its temperature decreases  \cite{gerlach1976mechanism,amelino2006black,PhysRevD.79.104009}.
On the other hand, our results in  Figs. (\ref{fig:1}) and (\ref{fig:2}) illustrate that increasing curvature, which is analogous to increasing the blackbody's mass, results in redshifted radiation whose properties closely resemble those of blackbody radiation at a lower temperature.
Therefore, the results obtained from our analog model of general relativity are consistent with the predictions of the Hawking-Bekenstein theory.

\begin{figure}
\centering
\includegraphics[width=0.8\columnwidth]{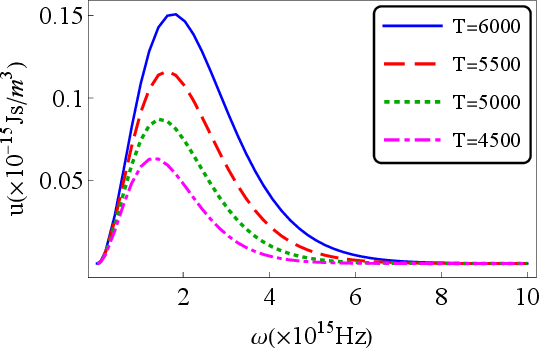}
\caption{ Energy density $u(\omega,T,\Lambda)$  versus $\omega$, for $\Lambda=0.3$; the thin-solid blue curve corresponds $T=6000K=0$, the dashed red curve to $T=5500K$, the dotted green curve to $T=5000K$, the dotted-dashed magenta curve to $T=4500K$
.}
 \label{fig:3}
\end{figure}

In Fig. \ref{fig:3} we have plotted the spectral distribution of energy in our blackbody radiation  with $\Lambda=0.3$  at different temperatures. As observed, in the presence of the curvature, the position of  the peak of the red-shifted radiation exhibits a strong temperature dependence. More precisely, increasing the temperature not only shifts the location of the peaks  towards  higher $\omega$, but also increases the height of each peak.
%

\subsection{ Curvature–dependent Stefan-Boltzmann’s Law of radiation} \label{sec3}
By integrating the curvature-dependent Plancks radiation law \eqref{12} over all frequencies, we obtain the generalized Stefan-Boltmann’s Law, which states that the total radiated energy $R(T)$
per unit surface area emitted by a black-body is proportional to $T^{4}$ \cite{johnson2012mathematical}.
By choosing  the dimensionless variable  $x\equiv \beta \hbar \omega$,  the generalized $\Lambda$–dependent energy density per unit frequency is given by:
 \begin{eqnarray}\label{14}
\frac{U}{v} &=& \int_{0}^{\infty} u(\omega)d\omega   \nonumber \\\\
 &=& \int_{0}^{\infty}  \frac{k^{4} T^{4}}{\pi^{2} \hbar^{3} c^{3}} x^{3}
  \frac{\sum_{n} (\Gamma n + \frac{\Lambda}{2} n^{2}) e^{-x(\Gamma n + \frac{\Lambda}{2} n^{2})}}{\sum_{n} e^{-x(\Gamma n + \frac{\Lambda}{2} n^{2})}}dx. \nonumber
\end{eqnarray}
In the limit of $\frac{\hbar \omega}{K T} \Lambda \ll 1$, Eq.\eqref{14} reduces to:
\begin{equation}\label{15}
\frac{U}{v}\simeq \frac{K^{4} T^{4}}{\pi^{2} \hbar^{3} c^{3}} \int_{0}^{\infty}  \frac{x^{3}}{e^{x}-1}\Big[1-\Lambda \frac{(x+1) e^{-x}+(x-1)}{(1-e^{-x})^{2}}\Big]dx.
\end{equation}
Consequently, the net rate of flow of radiation per unit area can be determined as follows:
 \begin{equation}\label{16}
R=\frac{1}{4}\frac{U}{v}c  = \sigma_{\Lambda} T^{4}.
\end{equation}
The proportionality of the total energy to the fourth power of the temperature is known as the Stefan-Boltzmann radiation law, formulated in 1879 \cite{loudon2000quantum}.
In our model, the curvature-dependent  Stefan-Boltmann’s constant is obtained as bellow:
\begin{equation}\label{17}
\sigma_{\lambda} \simeq \frac{1}{4}  \frac{k^{4} }{\pi^{2} \hbar^{3} c^{3}}\Big[\frac{ \pi^{4}}{15} - 18 \Lambda \zeta(3)\Big]  \quad    Wm^{-2}k^{-4}.
\end{equation}
It is evident that the curvature of the circle (representing the physical space) contributes to a reduction in the Stefan-Boltzmann constant.
%
\subsection{Curvature-dependent Rayleigh-Jeans’s energy density distribution}
To first order in the curvature $\Lambda$, the generalized energy density per unit frequency, Eq.~\eqref{15}, can be expressed as
\begin{equation}\label{18}
\tilde{u}(x)dx \simeq \frac{x^{3}}{e^{x}-1}\Big[1-\Lambda \frac{(x+1) e^{-x}+(x-1)}{(1-e^{-x})^{2}}\Big] dx ,
\end{equation}
where
$$\tilde{u}(x)dx \equiv \frac{\pi^{2} \hbar^{3} c^{3}}{k^{4} T^{4}}u(\omega)d\omega.$$
Here, we used change of variable: $x= \frac{\hbar \omega}{k T}$.

{\bf B1. Long wavelength limit :}
In the long wavelength limit: $x  \ll 1 $, we can approximate the  curvature–dependent Rayleigh-Jeans law as follows:
\begin{equation}\label{19}
\tilde{u}(x) \approx x^{2} \Big[1-\Lambda \frac{1-x}{x}\Big].
\end{equation}
As expected, in the flat limit  $\Lambda \rightarrow 0$, this expression
reduces to the standard classical Rayleigh-Jeans law, where $\tilde{u}(x) \approx x^{2} $ represents the normalized energy density~\cite{beale1996statistical}.

{\bf B2. Short wavelength limit :}

In the short wavelengths limit $ x  \gg 1 $, we find the generalized  Wien law as follows:
\begin{equation}\label{20}
\tilde{u}(x) \approx  x^{3} e^{-x} \Big[1+\Lambda (1-x)\Big].
\end{equation}
As expected, in the limit of $\Lambda \rightarrow 0$, this expression reduces to the standard Wien's law: $u(x) \approx x^{3} e^{-x}$ \cite{beale1996statistical}.
\begin{figure}
\centering
 \includegraphics[width=0.8\columnwidth]{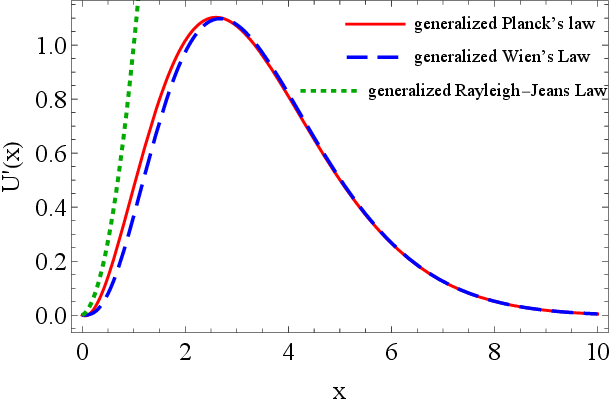}
\caption{ The spectral distribution of energy in the blackbody radiation; The thin-solid red curve corresponds the quantum theoretical formula of generalized Planck; the Dashed blue curve to the short-wavelength approximation of generalized Wien; the dotted green curve to the long-wavelength approximation of  generalized Rayleigh-Jeans law are also shown.}
 \label{fig:4}

\end{figure}
\section{ generalized Wien’s displacement law}\label{Wien law}
In this section, we consider curvature-dependent Wien’s displacement law from our curvature-dependent  blackbody radiation. The generalized spectral energy density  Eq.~\eqref{12}, as a function of wavelength $\lambda=\frac{2 \pi c}{\omega}$ and temperature $T$ can be written  as follows:
\begin{eqnarray}\label{21}
u(\lambda,T) &=& u(\omega,T)|\frac{d\omega}{d \lambda}| \simeq \frac{8 \pi h c}{ \lambda^{5}}  \frac{1}{e^{\frac{h c}{k\lambda T}}-1} \\\
&\times&\Big[1-\Lambda \frac{(\frac{h c}{k\lambda T}+1) e^{-\frac{h c}{k\lambda T}}+(\frac{h c}{k\lambda T}-1)}{(1-e^{-\frac{h c}{k\lambda T}})^{2}}\Big], \nonumber
\end{eqnarray}
where $|\frac{d\omega}{d \lambda}|$ represents the absolute value of the derivative, ensuring that the energy density remains positive~\cite{ma2009two}.
To determine Wien's displacement law, which describes the  $\lambda$-value of the peak of $u(\lambda, T)$ as a function of $T$ and $\lambda$ \cite{zettili2009quantum}, we must solve:
\begin{equation}\label{22}
[\frac{du(\lambda,T) }{d \lambda}]_{\lambda=\lambda_{max}}=0,
\end{equation}
where $\lambda_{max}$  is the wavelength at which the maximum of generalized Planck’s radiation formula occurs.

 By setting $y=\frac{h c}{k T \lambda}$, Eq. \eqref{22} can be written as:
 \begin{eqnarray}\label{23}
&& y^{2}e^{y}(1+4e^{y}+e^{y})\lambda_{c} +5 (-1+e^{y})^{2} \Big[-1+e^{y}(1+\lambda_{c})\Big]    \nonumber \\\
&& - y e^{y} (e^{y}-1) \Big[-1 +7 \lambda_{c} +e^{y}(1+7\lambda_{c})\Big]=0.
\end{eqnarray}
It is worth noting that in the flat limit, $\Lambda \rightarrow 0 $, it becomes
\begin{equation}\label{24}
ye^{y}-5 (-1+e^{y})=0,
\end{equation}
which is the standard version of Wein’s displacement law \cite{wien1896ueber,stavek2023hidden,kuhn1987black}.
However, due to the complexity of the final equation Eq. \eqref{23}, obtaining an analytical solution is challenging. Therefore, we resort to numerical methods to study its solutions.
In table \ref{tab2-2}, we present the numerical solutions of Eq. \eqref{23} and the corresponding $\lambda_{max}T(m K)$ values for various $\Lambda$-values. As evident from the results, an increase in curvature leads to an increase in the value of $\lambda_{max}T(m K)$.
 %
 \begin{table}[b]
  \centering
\caption{ $\lambda_{max}T$ for different $\Lambda$.}\label{tab2-2}
  \begin{tabular}{p{0.12\textwidth} p{0.15\textwidth} p{0.15\textwidth}}\hline
\hline
   $\Lambda$    & $y$   & $\lambda_{max}T(m  K)$ \\
  \hline
   $0.0$     &     $ 4.965$     &   $2.899 \times 10 ^{-3}$  \\
   $0.1$     &       $3.127$     &    $4.602 \times 10 ^{-3}$    \\
   $0.2$     &       $2.257$    &    $6.376 \times 10 ^{-3}$   \\
   $0.3$     &      $1.667$       &    $8.63 \times 10 ^{-3}$    \\
  \hline
\hline
\end{tabular}
\end{table}
 More precisely, it is seen that $\lambda_{max}T$ varies from value $2.899 \times 10 ^{-3} m K$ to  $8.63 \times 10 ^{-3} m K $,  as  the spatial curvature parameter $\Lambda$, varies  from $0$ to $0.3$.

\begin{figure}
\centering
 \includegraphics[width=0.8\columnwidth]{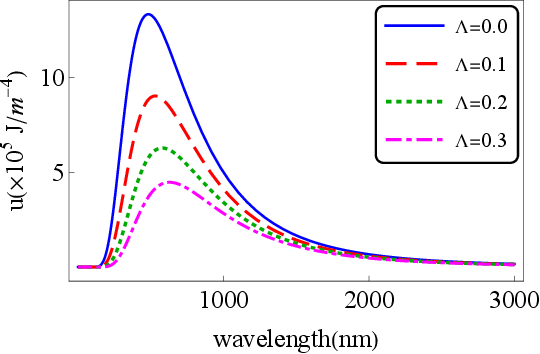}
\caption{ Energy density versus wavelength, $\lambda$,  $T=6000K$;  the thin-solid blue curve corresponds $\Lambda=0$, the dashed red curve to $\Lambda=0.1$, the dotted green curve to $\Lambda=0.2$, the dotted-dashed magenta curve to $\Lambda=0.3$.}
 \label{fig:5}
\end{figure}

In Fig.~\ref{fig:5} we have  plotted the energy density versus wavelength for a specific temperature $T=6000K$ and various curvature values. The emitted blackbody radiation exhibits a continuous intensity distribution with a maximum value that occurs at a curvature-dependent wavelength. Notably, the maximum radiation of a blackbody occurs at shorter wavelengths. Furthermore, it is clear that with increasing $\Lambda$, the peak of maximum radiation shifts towards longer wavelengths.

\begin{figure}
\centering
 \includegraphics[width=0.8\columnwidth]{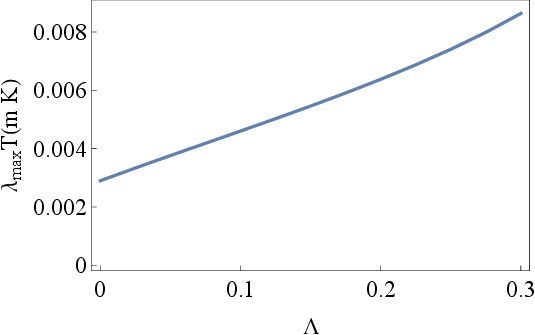}
\caption{ The variation of $\lambda_{max}T$ versus $\Lambda$.}
 \label{fig:6}
\end{figure}

Fig. \ref{fig:6} illustrates the variation of  $\lambda_{max}T$  with respect to the spatial curvature parameter $\Lambda$. Clearly, the value of
 $\lambda_{max}T$ increases with increasing  $\Lambda$.
This results in a shift of the peak of
the blackbody spectrum towards longer wavelengths (corresponding to lower frequencies) for a fixed temperature.
%
\section{SUMMARY AND CONCLUDING REMARKS}\label{SUMMARY}
We have presented a generalized Planck law for blackbody radiation, incorporating the effects of spatial curvature through an analog model of a simple quantum harmonic oscillator on a circle.
We  derived curvature-dependent energy density, Planck law, Stefan-Boltzmann law,  Rayleigh-Jeans law, and  Wien law of our  blackbody radiation, analyzing their behavior for various values of spatial curvature.
Our findings reveal that increasing spatial curvature leads to a decrease in both the height and width of the Planck radiation function, accompanied by a significant redshift in the peak frequency.
Furthermore, we found that the curvature of the circle (representing physical space) contributes to a reduction in the Stefan-Boltzmann constant while enhancing $\lambda_{max}T$ in  our curvature-dependent Wien’s displacement law.
These insights underline the critical role of spatial curvature in shaping blackbody radiation properties.

\section{Acknowledgements}
 The authors are grateful to support for this work by the Office of Graduate Studies and Research of Isfahan for their support, also wishes to thank the Shahrekord University for  their assistance.

\end{document}